%% file: hanhart_charm.tex
\documentclass[12pt]{article}
\usepackage{graphicx}

\def\pbnr{}
\def\speaker{Christoph Hanhart}

\def\title{Modelling low-mass resonances in
multi-body decays}
\def\affiliation{Institute for Advanced Simulation/Institut f\"ur Kernphysik \\
Forschungszentrum J\"ulich, J\"ulich, Germany}
\def\support{The workshop was supported by the University of Manchester, IPPP, STFC, and IOP}

\input charmmacros.tex

\begin{document}
\begin{titlepage}
\pubblock

\vfill
\Title{\title}
\vfill
\Author{\speaker\SupportedBy{\support}
}
\Address{\affiliation}
\vfill
\begin{Abstract}
Recent advances towards a model-independent description of hadronic
few-body final states, relevant for the extraction of direct CP violation from
the decays of heavy mesons, are presented.
\end{Abstract}
\vfill
\begin{Presented}
\venue
\end{Presented}
\vfill
\end{titlepage}
\def\thefootnote{\fnsymbol{footnote}}
\setcounter{footnote}{0}
%

\section{Introduction}

In order to explain the matter-antimatter asymmetry of the universe, an amount of CP violation
is necessary that exceeds that of the Standard Model (SM) by many orders of magnitude. Therefore,
CP violation qualifies as a very promising window to hunt for physics beyond the SM\footnote{In the context of these proceedings we also regard the QCD $\theta$--term, if it has a non--zero value, as physics beyond the
SM.}
of particle physics: it has to be somewhere and it has to exceed the SM predictions dramatically.
Here we focus on what it takes to extract possible CP violating signals in the quark sector, namely
from few--body decays of heavy mesons. For a discussion of another field of intense research,
 that of electric dipole moments of neutrons, protons and light nuclei, we refer to the literature (see Ref.~\cite{bira}
 and references therein).

If present, CP violation in the decay of heavy mesons will show up as a complex phase. As such, it can
only influence observables, if there are other amplitudes present that the CP violating amplitude can interfere
with. Therefore the decay of a heavy meson into three or more light mesons appears to provide an 
ideal environment for CP studies, for due to the presence of the large number of light meson resonances,
there is a lot of phase motion present in the amplitude. Not only should this phase motion make
the CP signals visible, because of its non--trivial distribution over the Dalitz plot, it at the same time
allows for a test of systematics. In addition it provides some sensitivity to the operator structure of
the CP violating source underlying the transition.

The mentioned benefit at the same time is the challenge: in order to be able to extract a CP phase from
the decay of a heavy meson into light quarks the final state interaction amongst the particles produced
must be controlled in a model-independent way---or at least with a controlled model dependence. 
It is the aim of this presentation to sketch the path to be followed to reach that goal as well as to
show where we stand.
It should be mentioned that there are also schemes that in principle allow for an  extraction of
CP signals from data directly~\cite{bediaga1,bediaga2}. However, we are convinced that using information
from theoretical analyses of the hadronic final--state interactions will increase the sensitivity to
CP violating signals significantly. This is especially important since direct CP violation in
$D$--meson decays, even if it exists, will certainly be very small~\cite{ikaros}.
To investigate these and related questions is part of the program of the informal
Les Nabis network~\cite{nabis}, which brings together physicists from theory and experiment
in heavy- and light-quark physics and aims at optimizing future Dalitz plot studies along
the lines sketched here.

\section{Theoretical tools}

Decay amplitudes must be consistent with, amongst others, the fundamental concepts of
analyticity and unitarity. In this section we will briefly outline how especially unitarity can
be used to constrain production amplitudes with a pair of low energy pions in the final
state, where here 'low energy' is to be understood as values of $Q^2{<}1$~GeV$^2$. Even in
the decay of heavy mesons there are  kinematic regions where this limit is realized.
In section \ref{further studies} some remarks will be given on how to extend the scheme
discussed here to larger values of $s$, as well as more than two particles.

\begin{figure}[t]
\begin{center}
\includegraphics[angle=0,width=.4\textwidth,height=!]{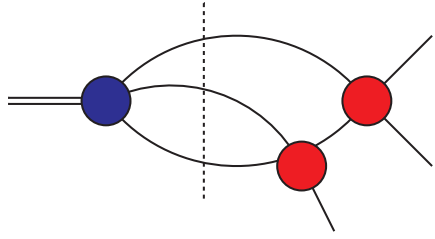}
\caption{Additional class of diagrams that contribute if more than
two strongly interacting particles are present. 
 \label{3body}}
\end{center}
\end{figure}

The discontinuity relation for a production amplitude reads\footnote{For simplicity written for two--body 
final states.}
\begin{equation}
{\rm Im}(F(Q^2)_i)=\sum_k T^*(Q^2)_{ik}\sigma(Q^2)_k
F(Q^2)_k \,\Theta(Q^2-M_k^2) \ ,
\label{imf}
\end{equation}
where the subindices $i$ and $k$ denote the final and intermediate channels, respectively, 
$\Theta(\ldots)$ is the Heaviside
step function, $M_k$ is the sum of the masses of the 
particles of channel $k$, $\sigma(Q^2)_k$ is the corresponding phase space and
$T_p(Q^2)_{ki}$ denotes the 
the scattering amplitude from channel $i$ to channel $k$.
If only a single channel contributes, Eq.~(\ref{imf}) encodes
the Watson theorem~\cite{watson}: since the left hand side is real valued, the right hand
tells us, that the phase of $F$ is necessarily identical to that of $T$.
Eq.~(\ref{imf}) holds for all production amplitudes, however, as soon as there are more
than two strongly interacting particle in the final state, the Watson theorem no longer
holds~\cite{caprini}, since rescattering diagrams of the kind shown
in Fig.~\ref{3body} introduce additional phases.

The pion vector form factor, $F_V$, is an ideal example to illustrate the implications 
and limitations of Eq.~(\ref{imf}). It is measured straightforwardly in $e^+e^-\to\pi^+\pi^-$,
but can also be extracted from decays of the $\tau$--lepton. 
It is defined via
\begin{equation}
\langle \pi^+(q_1)\pi^-(q_2)|J^\mu|0\rangle = (q_1-q_2)^\mu F_V(Q^2) \ ,
\label{fdef}
\end{equation}
where $Q=q_1+q_2$. The operator structure restricts the outgoing
pion pair into $p$--waves. Therefore, in the elastic regime
the scattering $T$--matrix
may be expressed via the corresponding phase shift $\delta_p(Q^2)$ as
\begin{equation}
T_p(Q^2) = \frac1{\sigma_\pi(Q^2)}\sin (\delta_p(Q^2))\exp(i\delta_p(Q^2)) \ .
\end{equation}
Below we use the phase shifts from the  analysis of Ref.~\cite{madrid}.

\begin{figure}
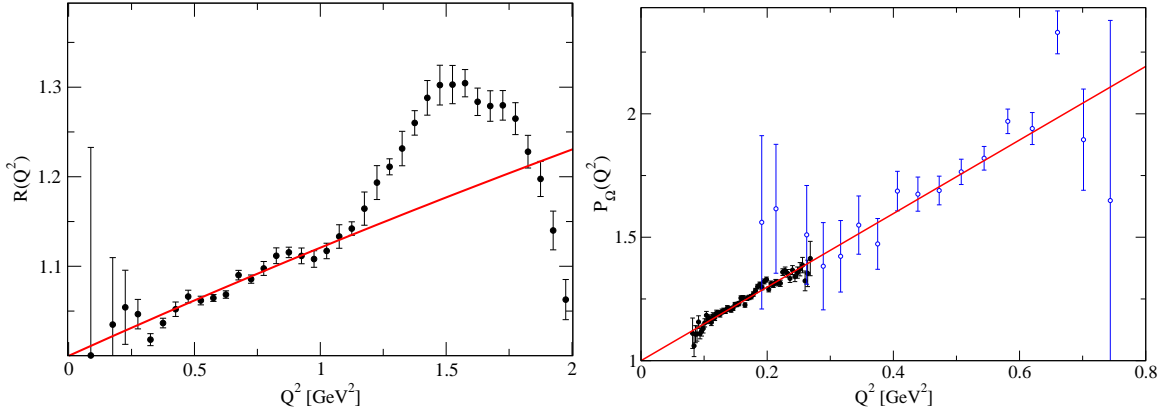

\includegraphics[angle=0,width=.5\textwidth,height=!]{tauFFoveromnes_zoom.eps}
\includegraphics[angle=0,width=.5\textwidth,height=!]{amplitudefit_K_etapoO.eps}
\caption{Left panel: Pion vector form factor, extracted from $\tau$ decays~\cite{belle}, divided
by the Omn\`es function, $\Omega(Q^2)$, defined in Eq.~(\ref{omnes}). Right
panel: Data on $\eta\to\pi\pi\gamma$~\cite{KLOE_2013}
  and  $\eta'\to\pi\pi\gamma$~\cite{Crystal_Barrel}, also divided
by $\Omega(Q^2)$. The figure is taken from Ref.~\cite{e2gg}}
\label{fig:overOmnes}
\end{figure}

\begin{figure}[t]
\begin{center}
\includegraphics*[width=0.5\linewidth]{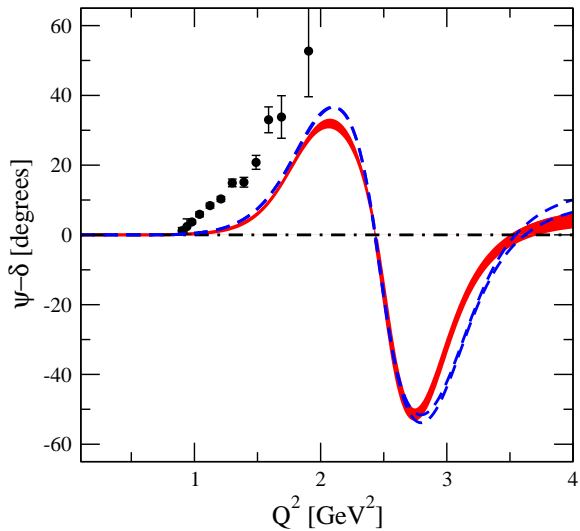}
\caption{Phase difference between $\psi$, the phase of the form factor,
and the scattering phase shift $\delta$. The data shown indicate the upper
bound of the phase shift difference presented in Ref.~\cite{Simon}.
 \label{phasediff}}
\end{center}
\end{figure}

If one assumes that the two-pion interactions are elastic up to infinite energies, the 
dispersion integral that emerges from Eq.~(\ref{imf}) can be solved analytically yielding
the celebrated Omn\`es function~\cite{Omnes:1958hv}
\begin{equation}
 \Omega(Q^2) 
    = \exp\left(\frac{Q^2}{\pi}\int_{4 m_\pi^2}^\infty \frac{ds}{s}\,\frac{\delta_p(s)}{s-Q^2-i\epsilon}\right)  \ .
\label{omnes}
\end{equation}
Since any function that is multiplied to  $F_V(Q^2)$ and that is real on the right-hand cut
does not spoil Eq.~(\ref{imf}),
one may write in general
\begin{equation}
F_V(Q^2)=R(Q^2)\Omega(Q^2) \ .
\label{rdef}
\end{equation}
In the left panel of Figure~\ref{fig:overOmnes} we show the $Q^2$ dependence of $R(Q^2)$.
As one can see, $R(Q^2)$ is perfectly linear 
for $Q^2<1$ GeV$^2$. For larger values of the $\pi\pi$ invariant mass squared one finds clear deviations
from linearity---in this case caused by the $\rho'$~\cite{newff}, the first radial excitation of the $\rho$-meson.
In the same energy range where the deviation from linearity is observed, also inelastic channels
become significant. This allows for a deviation from the Watson theorem as illustrated in Fig.~\ref{phasediff}:
the curves show the difference between $\delta$, the elastic $\pi\pi$ phase shift, and $\psi$, the
phase of the form factor, for two different variants of the model introduced in Ref.~\cite{newff}. The
data in this figure are the upper bounds for this phase--difference as derived from a data analysis
presented in Ref.~\cite{Simon}. 

The linearity of $R(Q^2)$ at low energies demonstrates very nicely 
that the bulk of the $Q^2$ dependence of the pion vector form factor
comes from the $\pi\pi$ final--state interactions, dominated by the $\rho$--meson. However, contrary
to the standard assumption of vector-meson-dominance models, this is not all: there is an additional
$Q^2$ dependence that must come from the vertex itself. It is reaction-dependent and needs to be
considered in any high-accuracy analysis of meson decays.

In order to derive Eq.~(\ref{rdef}) we only used the discontinuity equation, Eq.~(\ref{imf}), that constrains
the function $F$ on the right-hand cut. Thus, an analogous expression holds for
all amplitudes that have the same right-hand cut, as long as the left--hand cuts
are negligible. We may therefore write for the amplitude for
$\eta\to \pi\pi\gamma$
\begin{equation}
\mathcal{A}_{\pi \pi\gamma}^\eta(Q^2)=A_{\pi \pi\gamma}^\eta P_\Omega(Q^2)\Omega(Q^2) \ ,
\end{equation}
where, using $P_\Omega(0)=1$ and $\Omega(0)=1$ gives $\mathcal{A}_{\pi \pi\gamma}^\eta(0)
={A}_{\pi \pi\gamma}^\eta$. An analogous expression holds for the corresponding $\eta'$ decays.

In the right panel of Figure~\ref{fig:overOmnes} we show the $Q^2$ dependence of 
$P_\Omega(Q^2)$ for $\eta$ (solid symbols) as well as $\eta'$ (open symbols)  decays. 
The figure shows that $P_\Omega(Q^2)$  is linear within
the experimental uncertainties in the
full range kinematically accessible---although the data for $\eta'$ clearly call for improvement. 
The straight line in the figure is a fit to the $\eta$ data, which demonstrates that the 
slope of the $\eta'$ spectrum is consistent with that of the $\eta$.
The equality of the two slopes can be understood on the basis of chiral perturbation theory
in combination with large $N_c$ arguments~\cite{Stollenwerk}.
Note, while $F_V(Q^2)$ does not have a left-hand cut, the decay amplitudes for the
radiative decays of $\eta$ and $\eta'$  have one. Still, a linear
function in $Q^2$ is sufficient to parametrize the reaction specific 
energy dependence of the decay vertex.

It should be stressed that an important insight that comes from the investigations
presented is that it is not correct to parametrize the energy dependence of a
particular transition by a resonance term with the mass scale adjusted to the
reaction at hand, as it is often done in experimental analyses. What is demonstrated
here is that there are two sources of energy dependencies: on the one hand the final
state interactions. These are universal and driven largely by the resonance singularities.
On the other hand there is a reaction specific energy dependence of the particular
decay vertex. This function is a lot smoother than the final state interaction, but
cannot be ignored in high accuracy studies. 
It should also be noted that it is incorrect to add tree level diagrams to resonance contributions
in order to account for deviations from a resonance only picture as it is done, e.g., in 
Ref.~\cite{Carla_2010}, for this necessarily leads to a violation of the Watson theorem.

\section{Further studies}
\label{further studies}

In this section we briefly discuss a possible path to extend what was discussed in
the previous section to higher energies and what it takes to consider more than two
strongly interacting particles. 

We first focus on the extension of the formalism presented above
 to higher energies. Above $Q^2=1$~GeV$^2$ it is no longer justified
 to treat the $\pi\pi$ system as elastic~\cite{Simon} and inelastic 
 channels, in the $P$--wave most prominently 4$\pi$ and $\pi\omega$, need to be included.
 In this energy range the Watson theorem does not hold anymore and
 Eq.~(\ref{omnes}) can no longer be applied. If one uses the simplifying assumption
 that the coupling of inelastic channels to the $\pi\pi$--system is of short range,
 then a system of algebraic equations can be derived that allows for a convenient
 fit of the available data over a large energy range with the fit parameters
 largely given by couplings and mass parameters of the higher resonances; by construction the
 formalism maps on a treatment in terms of the Omn\`es function at low energies
 smoothly~\cite{newff}. For this energy range, elastic $p$--wave $\pi\pi$ phase shifts must be used as input.
 
 The same kind of approach can in principle also be used for partial waves other 
 than $p$--waves. For the isoscalar $s$--waves using dispersion theory to constrain the amplitudes
 is of particular importance, since their behavior deviates significantly from that of a
 Breit-Wigner~\cite{juergulf,susanulf}.
 However, the problem  arises that the first
 inelastic channel sets in very prominently at $Q^2=4m_K^2$. It is therefore not clear
 what to use as input phases, for elastic phase shifts are needed if one wants
 to continue using Eq.~(\ref{omnes}) to describe the low energy regime. Alternatively
 one could use numerical solutions of the coupled--channel Omn\`es equation in
 this energy regime, which are available numerically~\cite{DGL90,Moussallam99,Descotes,HDKM,ours}. 
The additional problem here is that direct data for the scalar from factors are not available,
since there is no scalar probe. As a consequence of this
the scalar $\pi\pi$ interactions appear only within subsystems of strongly interacting
few particle systems and all issues raised above apply. 

\begin{figure}[t]
\begin{center}
\includegraphics*[width=0.6\linewidth]{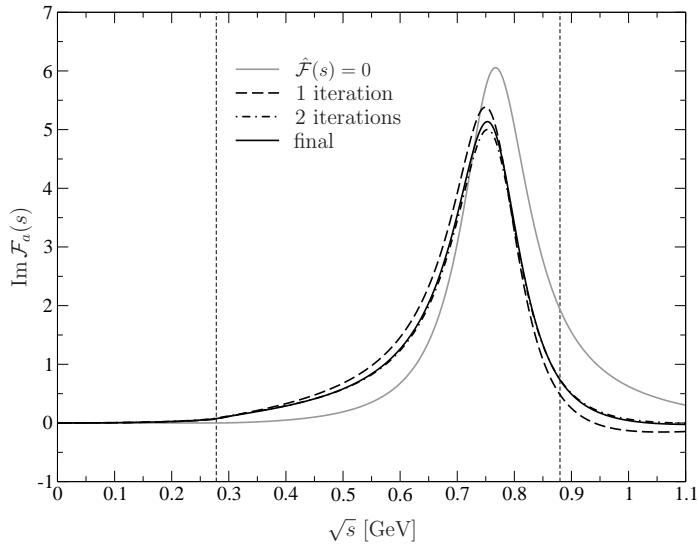}
\caption{Effect of three--body interactions on the imaginary part of
the decay amplitude $\phi\to 3\pi$. The case $\hat{\cal F}=0$ refers
to consideration of two--body interactions only. The other two curves
correspond to an iterative solution of full the set of equations.
 \label{3body_effect}}
\end{center}
\end{figure}

In order to fully consistently treat a system of three strongly interacting particles, the
whole formalism must be generalized, for also diagrams of the
kind shown in Fig.~\ref{3body} need to be included in the dispersive treatment. 
The relevant equations were derived
already in 1960~\cite{3b_method} and first applied to $\omega\to 3\pi$ in Ref.~\cite{3b_omegaold}.
A convenient strategy to solve the equations is given in Ref.~\cite{3b_technique}---especially here the notion of the hat-functions was introduced that allow for the inclusion
of interactions within subsystems.
The most recent analysis of this kind is presented 
 in Ref.~\cite{franz},
where the full 3$\pi$ interactions are included in a dispersive analysis of $\omega\to 3\pi$
and $\phi\to 3\pi$. Since pion pair interactions can only happen
in odd partial waves, the formalism simplifies considerably, especially since $F$--waves or
higher turner out to be negligible. 
The results for the imaginary part of the $\phi$ decay amplitude is shown in Fig.~\ref{3body_effect}.
The case $\hat{\cal F}=0$ refers to omission of rescattering effects.
Then the equations were solved iteratively. Although the dispersive treatment can not
be mapped directly onto Feynman diagrams, still an increasing number of iterative steps
can be viewed as the consideration of an increasing level of complexity.
It is therefore found that rescattering effects indeed lead to a significant distortion
of the line shape as well as a modification of the phase motion associated to the final state
interaction---note that even the peak position is shifted via the rescattering effects, although
the location of the pole in the $\pi\pi$--subsystem remains unchanged. This is a clear illustration
that any kind of Breit-Wigner parametrization should not be used in an analysis of high-accuracy data. 

Unfortunately the effect of the third particle depends not only on the partial
waves involved but also on the total energy available for the reactions. It is therefore 
not possible to conclude from the findings of Ref.~\cite{franz} on the general importance of rescattering
effects in the decay of heavy mesons. However, it is clear that a systematic understanding also
of these effects is necessary in order to allow for a high accuracy description of 
three-- or more--body decays.

\section{Outlook}

Three-- and more--body decays of $D$--mesons are good candidates to look
for CP--violations beyond the Standard Model. Not only is the signal expected
from the SM tiny, the three--body nature of the final state allows one to control
the systematics of the analysis and may even reveal valuable information on the
mechanism underlying the CP--violation beyond the SM, once it is discovered.

However, such an analysis necessitates a high-accuracy control over the hadronic
final state interactions that allows for a systematic control of the uncertainties. 
As outlined here this is in principle possible using dispersion theory. Although the
fundamental equations are known since long, their full implementation is still
quite demanding, especially when it comes to three or 
more strongly interacting particles. Especially the inclusion of inelastic channels
requires additional work. The corresponding research is under way.

\vspace{0.2cm}

{\bf Acknowledgments}

I am grateful to Bastian Kubis and Franz Niecknig for helpful discussions
and careful reading of the manuscript.
This work is supported by the DFG and the NSFC (Sino-German CRC 110 ``Symmetries
and the Emergence of Structure in QCD'') and by the
EU Integrated Infrastructure Initiative HadronPhysics3.

\end{document}

%% file: charmmacros.tex
\newcommand\pubnumber{\pbnr}
\newcommand\pubdate{\today}
\def\Title#1{\begin{center} {\Large #1 } \end{center}}
\def\Author#1{\begin{center}{ \sc #1} \end{center}}

\newcommand{\OnBehalf}[1]{\sbox0{#1}\ifdim\wd0=0pt
        {}
	\else
	{\\on behalf of #1}
	\fi}
\newcommand{\SupportedBy}[1]{\sbox0{#1}\ifdim\wd0=0pt
        {}
	\else
	{\footnote{#1}}
	\fi}
\def\Address#1{\begin{center}{ \it #1} \end{center}}

\newcommand\pubblock{\includegraphics[width=5cm]{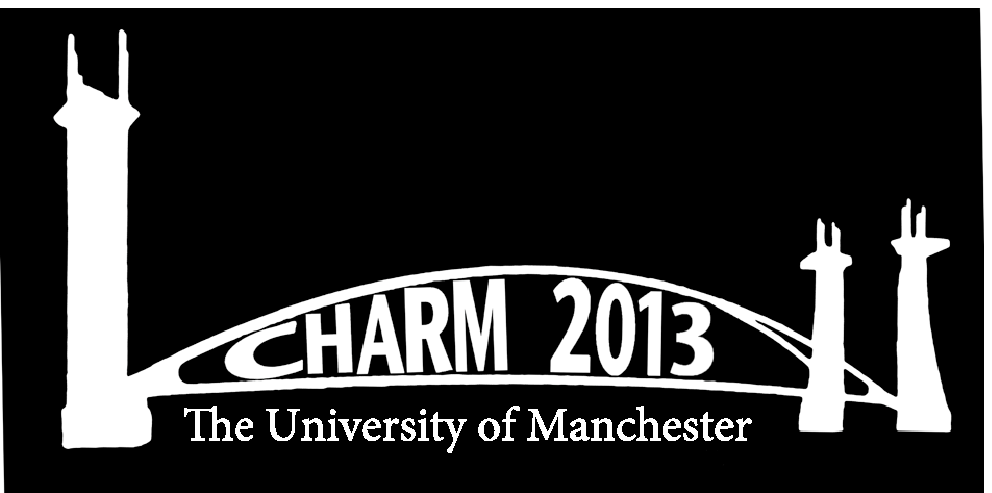}\hfill{\begin{tabular}{l} \pubnumber\\
         \pubdate  \end{tabular}}}
\newenvironment{Abstract}{\begin{quotation}  }{\end{quotation}}
\newenvironment{Presented}{\begin{quotation} \begin{center} 
             PRESENTED AT\end{center}\bigskip 
      \begin{center}\begin{large}}{\end{large}\end{center} \end{quotation}}

\def\venue{The 6$^{th}$ International Workshop on Charm Physics\\
(CHARM 2013)\\
Manchester, UK,  31 August -- 4 September, 2013}

\def\beq{\begin{equation}}
\def\eeq#1{\label{#1}\end{equation}}
\def\eeqn{\end{equation}}

\def\beqa{\begin{eqnarray}}
\def\eeqa#1{\label{#1}\end{eqnarray}}
\def\eeqan{\end{eqnarray}}

\let\bar=\overbar

\def\Dslash{\not{\hbox{\kern-4pt $D$}}}
\def\dslash{\not{\hbox{\kern-2pt $\del$}}}

\def\msb{{\bar{\ssstyle M \kern -1pt S}}}

